\documentstyle[12pt]{article}
\def\theequation{\arabic{section}.\arabic{equation}}
\begin{document}
\renewcommand{\theequation}{\thesection.\arabic{equation}}

\begin{center}
{{\Large{The $sl(2n|2n)^{(1)}$ Super-Toda Lattices and}
 \\ {the Heavenly Equations as Continuum Limit}}}
\end{center}
\vskip0.4cm
\begin{centerline}
{\large{Zhanna Kuznetsova\footnote{{\em E-mail: zhanna@cbpf.br}}, 
Ziemowit Popowicz\footnote{{\em E-mail: ziemek@ift.uni.wroc.pl}}
and Francesco Toppan\footnote{{\em E-mail: toppan@cbpf.br}}
}}
\end{centerline}
\small{
\begin{center}
${}^1${\it Dep. de F\'{\i}sica, Universidade Estadual de Londrina,}\\
{\it Caixa Postal 6001, Londrina (PR), Brazil}
\end{center}
\begin{center}
${}^2${\it Institute for Theoretical Physics, University of
Wroc{\l}aw,}
\\ {\it 50-204 Wroc{\l}aw, pl. Maxa Borna 9, Poland}
\end{center}
\begin{center}
${}^3${\it CBPF, Rua Dr. Xavier Sigaud 150,}\\ {\it cep 22290-180 Rio de Janeiro (RJ), Brazil}
\end{center}
}
\vskip1cm
\begin{abstract}
The $n\rightarrow\infty$ continuum limit of super-Toda models associated with
the affine $sl(2n|2n)^{(1)}$ (super)algebra series produces $(2+1)$-dimensional integrable equations in the ${\bf S}^{1}\times {\bf R}^2$ spacetimes. The equations of motion of the
(super)Toda hierarchies depend not only on the chosen (super)algebras but also on the specific
presentation of their Cartan matrices. Four distinct series of integrable hierarchies
in relation with symmetric-versus-antisymmetric, null-versus-nonnull presentations of the corresponding Cartan matrices are investigated. In the continuum limit we derive four classes of integrable equations of heavenly type, generalizing
the results previously obtained in the literature.
The systems are manifestly $N=1$ supersymmetric and, for specific choices of the Cartan matrix preserving the complex structure,
admit a hidden $N=2$ supersymmetry. The coset reduction of the (super)-heavenly equation to the
${\bf I}\times{\bf R}^{(2)}=({\bf S}^{1}/{\bf Z}_2)\times {\bf R}^2$ spacetime
(with ${\bf I}$ a line segment) is illustrated.
Finally, integrable $N=2,4$ supersymmetrically extended models in $(1+1)$ dimensions are
constructed through dimensional reduction of the previous systems. 
\end{abstract}
\vskip3cm
\vfill{CBPF-NF-009/05}
\thispagestyle{empty}\vfill

\section{Introduction}
The class of the so-called ``heavenly equations" has been introduced by Pleba\'nski \cite{ple}
to describe the solutions of the self-dual Einstein gravity. In this context the heavenly equations
have been widely analyzed, their relation with the $SU(\infty)$ Toda fields equations being pointed
out (see e.g. \cite{tod} and the references therein).
\par
The integrability properties of such equations were investigated in a series of papers, see \cite{int}, while,
in the beginning of the nineties, it was further proven \cite{ov} that the first heavenly equation describes the string theory with local $N=2$ supersymmetry (for a recent review containing an updated list of references see \cite{gc}).
The connection with superstrings provides a strong physical motivation for the construction of the integrable supersymmetric extensions of the heavenly equations (to be recovered from superalgebraic data).\par
For what concerns such supersymmetric extensions, the present state of the art is as follows. Saveliev and Sorba \cite{ss} first derived the $(2+1)$-dimensional $N=1$ supersymmetric heavenly equation as
a continuum limit (for $n\rightarrow \infty$) of a discretized super-Toda system based on the finite $sl(n|n+1)$
superalgebra series. Discretized super-Toda systems were further investigated in several papers \cite{ls1}. 
The systems analyzed in \cite{ls2} admit a hidden $N=2$ supersymmetry. In \cite{pt} it was shown, following \cite{it}, how to exploit
the complex structure of $sl(n|n+1)$ to encode data in a way to construct an integrable, manifestly $N=2$ supersymmetric, Toda lattice admitting an $N=2$ superheavenly equation in its continuum limit
(the superheavenly equation of reference \cite{ss} being recovered as a special reduction).\par
In this paper we address two separate issues and present their solutions. In this way we are able to enlarge the
class of integrable supersymmetric Toda lattices (as well as their continuum limit) produced in the literature
with the
introduction of new sets of integrable equations.\par
We recall at first the interpretation of the dots of the (super)algebra Dynkin diagrams as discretized
positions of the (super)Toda lattice which label, in the continuum limit, an extra-dimension. For the class of
(super)Toda lattices based on {\em finite} (super)Lie algebras, as the ones studied in \cite{{ss},{pt}}, the discretized
extra dimension corresponds to the half-line ${\bf R}_+=[0,+\infty]$. The continuum limit of the superToda lattice
therefore produces a $(2+1)$-dimensional integrable system based on either ${\bf R}_+\times {\bf R}^{(2,0)}$ or
${\bf R}_+\times {\bf R}^{(1,1)}$ (according to whether the (super)Toda lattice is formulated as a relativistic two-dimensional Euclidean or Minkowskian theory). On the other hand,
due to ciclicity of the affine (super)Lie algebra Dynkin diagrams introduced below, their dots produce a discretization of the ${\bf S}^{1}$ circle. A discretized (super)Toda lattice based on, e.g., the $sl(2n|2n)^{(1)}$ affine superalgebra series,
see \cite{kac} and \cite{dic}, produces in the continuum limit a class of $(2+1)$ integrable systems whose base spacetimes is either ${\bf S}^1\times {\bf R}^{(2,0)}$ or ${\bf S}^1\times {\bf R}^{(1,1)}$.
It can be further proven that a coset construction 
${\bf I}={\bf S}^{1}/{\bf Z}_2$ allows to express the continuum limit in the  
${\bf I}\times {\bf R}^{(2,0)}$ or ${\bf I}\times {\bf R}^{(1,1)}$ spacetimes,
where ${\bf I}$ is a closed interval which, without loss of generality, can be chosen to be given by ${\bf I}= [0,1]$.\par
In the present paper, unlike the previous works in the literature based on {\em finite} superalgebras, we construct  superToda lattices associated with the $sl(2n|2n)^{(1)}$ {\em affine} superalgebras
(the reason for us to work with this special class of $n$-parametric affine superalgebras is the presence, see \cite{dic}, of a complex structure; due to that, and depending on the given chosen construction used in the following, in special cases we are able to obtain $N=2$ supersymmetric systems), producing in the 
 $n\rightarrow \infty$ continuum limit the ${\bf S}^1\times {\bf R}^{2}$ systems previously discussed.
\par
It should be reminded that a (super)Lax pair of a $2$-dimensional (super)Toda system only requires the algebraic data
contained in the given, associated, Cartan matrix. The second main issue addressed in this paper
concerns the fact that, however, the  (super)Toda model is {\em not} uniquely specified by
its associated Cartan matrix. An algebraically equivalent, but a different presentation for the Cartan matrix,
actually produces a different type of Toda model. This feature often passes unnoticed and its importance not recognized,
implying that whole classes of integrable systems are not correctly identified and are instead disregarded.\par
For super-algebras it is already well-known that we need to work with the specific presentation of Dynkin diagrams realized by {\em all} fermionic simple roots (whenever this is indeed possible, following the superalgebras classification,
see \cite{kac} and \cite{dic}), in order to have a linearly implemented,
manifest supersymmetric Toda system of equations. The other Dynkin superalgebra presentations, according to \cite{st}, produce spontaneously broken, non-linearly realized, supersymmetric Toda models.\par
However, a given presentation of the Cartan matrix is already important for the derivation of Toda models 
associated with bosonic Lie algebras (as well as for super-Toda models 
derived from Cartan matrices associated with {\em all} fermionic simple roots Dynkin diagrams). It is known \cite{kac2}
that, without loss of generality, the Cartan matrix can be chosen, e.g., either symmetric or antisymmetric.
The derivation of the bosonic heavenly equation as the $n\rightarrow \infty$ continuum limit of
the $sl(n)$ Toda lattice {\em requires} the symmetric presentation of the $sl(n)$ Cartan matrix (given by 
$A_{ij} = 2\delta_{ij} -\delta_{i,j+1} -\delta_{i,j-1}$) which makes it to correspond to a discretization of the second-order derivative entering the bosonic heavenly equation \cite{int}).\par
For what concerns the $sl(2n|2n)^{(1)}$ affine superalgebra series, see e.g. \cite{dic}, we can work within a
purely fermionic simple roots cyclic Dynkin diagram. Four main classes of presentations for the associated Cartan matrices
$A_{ij}$
can be considered. They can be obtained by combining\\
{\em i}) the symmetric presentation for the Cartan matrix or,\\
{\em ii}) its antisymmetric counterpart, as well as,\\
{\em a}) the ``null" presentation
(i.e. for any $i$, $\sum_j A_{ij}=0$ and, similarly, for any $j$, $\sum_i A_{ij}= 0$) or,\\
{\em b})  the ``non-null" presentation
(such that $\sum_j A_{ij}\neq0$, $\sum_iA_{ij}\neq 0$, $|\sum_jA_{ij}|= |\sum_iA_{ij}|=2 $). \par
For each given $n$, as well as in the $n\rightarrow\infty$ continuum limit, four corresponding 
classes of supersymmetric
systems of integrable equations, labeled by ${\em ia}$), ${\em i b}$),
${\em ii a}$) and ${\em ii b}$) respectively, are produced accordingly.\par
For us it is convenient to work with the $N=1$ superfield formalism of \cite{tz}, rather than the $N=2$ formalism
introduced in \cite{it} and also used in \cite{pt}. The reason is that the $N=1$ formalism is more general. Due to the existence of a complex structure, if the latter is preserved by the specific presentation of the Cartan matrix, the
associated Toda system automatically admits a hidden $N=2$ supersymmetry. On the other hand, the $N=2$ formalism does not allow us to reproduce the super-Toda systems which possess only an $N=1$ supersymmetry. In the following we construct a whole new class of $N=1$ super-Toda models.\par
The Toda models analyzed in this paper (for any value of $n=1,2\ldots$ and in the limit $n\rightarrow \infty$, as well
as for each one of the four presentations of the Cartan matrices) are all given by a coupled non-linear system
of two $N=1$ superfields, plus two extra superfields satisfying equations in the background produced by the
first two coupled superfields.\par
The scheme of the paper is as follows. In the next Section we introduce, following \cite{tz}, the $N=1$ Lax pair
formulation for super-Todas and apply it to derive the four models associated with the $sl(2|2)^{(1)}$ superalgebra.
In Section ${\bf 3}$ we extend these results to any $n$, producing the $sl(2n|2n)^{(1)}$ super-Toda lattices and
further deriving their continuum limit (corresponding to the four classes of super-heavenly equations). In Section
${\bf 4}$ we show how to reduce a (super)-heavenly type of equation from ${\bf S}^1\times {\bf R}^2$ to its coset
${\bf I}\times {\bf R}^2$, with ${\bf I}= {\bf S}^1/{\bf Z}_2$. In Section ${\bf 5}$ we discuss the $(1+1)$ dimensional
reductions of the previous systems, producing super-hydrodynamical type of equations with $N=1,2,4$ supersymmetries.
In the Conclusions we address some open problems concerning the construction of $N>2$, $N$-extended supersymmetric
two-dimensional Toda models and $(2+1)$-dimensional heavenly types of equation.   
 
\section{The $sl(2|2)^{(1)}$ super-Toda models.}

Let us introduce at first, following \cite{tz}, the manifest $N=1$ formalism which will be employed throughout this paper.
The fermionic derivatives are given by
\begin{eqnarray}
D_\pm &=& \frac{\partial}{\partial\theta_\pm}
+{\theta}_\pm \partial_\pm,
\end{eqnarray}
where $\theta_\pm$ are fermionic Grassmann coordinates, while $x_\pm$ are the light-cone coordinates
($\partial_\pm=\frac{\partial}{\partial x_\pm}$) s.t. $x_\pm = x\pm t$ in the Minkowskian
case and $x_\pm=x\pm it$ in the Euclidean (in the latter case $x_+={x_-}^\ast$).  \par
The $N=1$ supersymmetric Lax pairs are given by
\begin{eqnarray}
L_+ &=& D_+\Phi +e^{\Phi}F_+e^{-\Phi},\nonumber\\
L_- &=& -D_-\Phi+ e^{-\Phi}F_-e^{\Phi},
\end{eqnarray}
where
\begin{eqnarray}
\Phi &=& \frac{1}{2}\sum_j \Phi_jH_j,\nonumber\\
F_+&=& \sum_j F_{+j},\nonumber\\
F_-&=&\sum_jF_{-j}.
\end{eqnarray}
In the above formula ``$\Phi_j$"' denotes a set of ($N=1$) bosonic superfields; the sums over $j$ are restricted to,
respectively, the Cartan generators and the simple (positive and negative) fermionic roots.
For our purposes we only need to know the following algebraic relations between Cartan generators and fermionic simple roots,
given by
\begin{eqnarray}
\relax [H_i, F_{\pm j}] &=& \pm A_{ij}F_{\pm j}, \nonumber\\
\{F_i, F_{-j} \} &=& \delta_{ij} H_j,
\end{eqnarray}
where $A_{ij}$ denotes the Cartan matrix.\par
The zero-curvature equation, given by,
\begin{eqnarray}
\{ D_++L_+,D_-+L_-\}&=&0
\end{eqnarray}
reproduces the set of super-Toda equations (the sum over repeated indices is understood)
\begin{eqnarray}
D_+D_-\Phi_j&=& \exp{(\Phi_i A_{ij})}.
\end{eqnarray} 
Let us specialize now the above system to the four specific presentations for the $sl(2|2)^{(1)}$ 
Cartan matrix discussed in the introduction. They are respectively given by \cite{kac}, \cite{dic}
\begin{eqnarray}
{\em ia})&\equiv  
&
\left(\begin{tabular}{cccc}
  $0$&$1$&$0$&$-1$\\ 
  $1$&$0$&$-1$&$0$\\
  $0$&$-1$&$0$&$1$\\
  $-1$&$0$&$1$&$0$
   \end{tabular}\right)
\end{eqnarray}
\begin{eqnarray}
{\em ib})&\equiv  
&
\left(\begin{tabular}{cccc}
  $0$&$1$&$0$&$1$\\ 
  $1$&$0$&$1$&$0$\\
  $0$&$1$&$0$&$1$\\
  $1$&$0$&$1$&$0$
   \end{tabular}\right)
\end{eqnarray}
\begin{eqnarray}
{\em iia})&\equiv  
&
\left(\begin{tabular}{cccc}
  $0$&$1$&$0$&$-1$\\ 
  $-1$&$0$&$1$&$0$\\
  $0$&$-1$&$0$&$1$\\
  $1$&$0$&$-1$&$0$
   \end{tabular}\right)
\end{eqnarray}
\begin{eqnarray}
{\em iib})&\equiv  
&
\left(\begin{tabular}{cccc}
  $0$&$1$&$0$&$1$\\ 
  $-1$&$0$&$-1$&$0$\\
  $0$&$1$&$0$&$1$\\
  $-1$&$0$&$-1$&$0$
   \end{tabular}\right)
\end{eqnarray}
It is perhaps worth showing the explicit generators redefinitions which allow to connect
the different presentations of the Cartan matrices. We have\\
{\em 1}) The {\em ia})$\rightarrow$ {\em ib}) transformation corresponds to the mappings \par
$H_3\rightarrow -H_3$, $H_4\rightarrow -H_4$, $F_{\pm 3}\rightarrow i F_{\mp 3}$, $F_{\pm 4}\rightarrow
iF_{\mp 4}$ (the other generators are left unchanged),\\
{\em 2}) the {\em ib}) $\rightarrow$ {\em iia}) transformation is obtained through \par
$H_2\rightarrow -H_2$, $H_3\rightarrow -H_3$, $F_{\pm 3}\rightarrow i F_{\mp 3}$, $F_{\pm 2}\rightarrow
iF_{\pm 2}$, $F_{\pm 4} \rightarrow F_{\mp 4}$ (the other generators are unchanged),\\
{\em 3}) the {\em ib}) $\rightarrow$ {\em iib}) transformation corresponds to \par
$H_2\rightarrow -H_2$, $H_4\rightarrow -H_4$, $F_{\pm 2}\rightarrow i F_{\pm 2}$, $F_{\pm 4}\rightarrow
iF_{\pm 4}$ (the other generators are unchanged).\par
Before writing the four sets of super-Toda equations corresponding to the $sl(2|2)^{(1)}$ affine superalgebra,
let us set at first
\begin{eqnarray}
\Phi_{\pm} &=&\frac{1}{2} (\Phi_3\pm\Phi_1),\nonumber\\
\Lambda_{\pm} &=&\frac{1}{2}(\Phi_4\pm \Phi_2)
\end{eqnarray}
We have
\\
{\em Case ia})
This presentation of the $sl(2|2)^{(1)}$ Cartan matrix was employed in \cite{it}. The Cartan matrix 
is degenerate (with rank two) and the system is not conformally invariant. In \cite{it}, in order
to remove the degeneration of the Cartan matrix, the $sl(2|2)^{(1)}$ superalgebra was extended with the
addition of two extra Cartan generators. The system discussed in \cite{it} was based on
the conformally affine approach pionereed in \cite{bb} and proved to be a conformal extension of the
$N=2$ sinh-Gordon equation with a spontaneous breaking of the conformal symmetry. For the non-extended 
$sl(2|2)^{(1)}$ affine superalgebra we get here
\begin{eqnarray}
D_+D_-\Phi_-&=& \sinh(2\Lambda_-),\nonumber\\
D_+D_-\Lambda_-&=&\sinh(2\Phi_-),
\end{eqnarray}
together with
\begin{eqnarray}
D_+D_-\Phi_+&=& \cosh(2\Lambda_-),\nonumber \\
D_+D_-\Lambda_+&=&\cosh(2\Phi_-).
\end{eqnarray}
This system of equations corresponds to two coupled sinh-Gordon equations plus two background equations.
The complexification of the coupled sinh-Gordon superfields induces a hidden $N=2$ supersymmetry. Indeed,
the second supersymmetry closing the $N=2$ invariance is explicitly given by the transformations 
(associated with the light-cone coordinates $x+$, $x_-$, respectively)
\begin{eqnarray}
\delta_+ \Phi_\pm = i\epsilon D_+\Phi_\pm, &\quad & \delta_+\Lambda_\pm = -i\epsilon
D_+\Lambda_\pm,\nonumber\\
\delta_- \Phi_\pm = i\epsilon D_-\Phi_\pm, &\quad & \delta_-\Lambda_\pm = -i\epsilon
D_-\Lambda_\pm.
\end{eqnarray}
A consistent $N=1$ reduction corresponds to identify $\Phi_\pm =\Lambda _\pm$ 
(in the minimal case $\Phi_\pm$
can be assumed to be a {\em real} superfield), while a consistent
$N=2$ reduction corresponds to set
$\Phi_\pm ={\Lambda_\pm}^\ast$ for a {\em complex} superfield $\Phi_\pm$. \\
\\
{\em Case ib}) We get
\begin{eqnarray}
D_+D_-\Phi_+&=& \exp(2\Lambda_+),\nonumber\\
D_+D_-\Lambda_+&=&\exp(2\Phi_+),
\end{eqnarray}
together with
\begin{eqnarray}
D_+D_-\Phi_-&=& 0,\nonumber \\
D_+D_-\Lambda_-&=&0,
\end{eqnarray}
This system of equations corresponds to two coupled (super)Liouville equations plus two free equations.
As before, the complexification of the coupled Liouville superfields gives an $N=2$ supersymmetry.
We have a consistent $N=1$ reduction, given by $\Phi_\pm =\Lambda _\pm$, and a consistent $N=2$
reduction for
$\Phi_\pm ={\Lambda_\pm}^\ast$.\\
\\
{\em Case iia})
\begin{eqnarray}
D_+D_-\Phi_-&=& \sinh(-2\Lambda_-),\nonumber\\
D_+D_-\Lambda_-&=&\sinh(2\Phi_-),
\end{eqnarray}
together with
\begin{eqnarray}
D_+D_-\Phi_+&=& \cosh(2\Lambda_-),\nonumber\\
D_+D_-\Lambda_+&=&\cosh(2\Phi_-),
\end{eqnarray}
This system of equations corresponds to two coupled sinh-Gordon equations with ``wrong" sign plus two background equations.
The ``wrong" sign means that $\Phi_\pm =\Lambda _\pm$ is {\em not} a consistent reduction and that the
complexification of the above coupled Sinh-Gordon superfields {\em does not} produce an $N=2$ supersymmetry.\\
\\
{\em iib})
\begin{eqnarray}
D_+D_-\Phi_+&=& \exp(-2\Lambda_+),\nonumber \\
D_+D_-\Lambda_+&=&\exp(2\Phi_+),
\end{eqnarray}
together with
\begin{eqnarray}
D_+D_-\Phi_-&=& 0,\nonumber\\
D_+D_-\Lambda_-&=&0,
\end{eqnarray}
This system of equations corresponds to two coupled Liouville equations with ``wrong" sign plus two free equations.
As before, the ``wrong" sign means that $\Phi_\pm =\Lambda _\pm$ is {\em not} a consistent reduction and that the
complexification of the above coupled Liouville superfields {\em does not} produce an $N=2$ supersymmetry.

\section{The affine $sl(2n|2n)^{(1)}$ super-Toda lattices and their continuum limit.}

Let us now discuss the different classes of $sl(2n|2n)^{(1)}$ super-Toda lattices and their continuum limits.
We work with the fermionic presentation of the simple roots, the Dynkin diagram being cyclic and admitting
$4n$ dots. The four associated presentations of the Cartan matrices are given, for $k,l=1,\ldots , 4n$, by
\begin{eqnarray}&\begin{array}{lcc}
{ia)} & A_{kl} =& -\delta_{1,k}\delta_{l,4n}-\delta_{k,4n}\delta_{l,1}+ (-1)^k\delta_{k,l+1}-(-1)^k\delta_{l,k+1},\nonumber\\
{ib)} & A_{kl} =& \delta_{1,k}\delta_{l,4n}+\delta_{k,4n}\delta_{l,1}+ \delta_{k,l+1}+\delta_{l,k+1},\nonumber\\
{iia)}&  A_{kl} =& -\delta_{1,k}\delta_{l,4n}+\delta_{k,4n}\delta_{l,1}-\delta_{k,l+1}+\delta_{l,k+1},\nonumber\\
{iib)}& A_{kl} =& \delta_{1,k}\delta_{l,4n}-\delta_{k,4n}\delta_{l,1}- (-1)^k\delta_{k,l+1}-(-1)^k\delta_{l,k+1}.
\end{array}&\\
\end{eqnarray}
For our purposes it is convenient to group the superfields as follows, where $i$ plays the role of
a discretized extra-time variable,
\begin{eqnarray}
B_i &=& \Phi_{2i}-\Phi_{2i-2},\nonumber\\
C_i&=& \Phi_{2i}+\Phi_{2i-2},\nonumber\\
E_i&=& \Phi_{2i+1}-\Phi_{2i-1},\nonumber\\
F_i&=& \Phi_{2i+1}+\Phi_{2i-1}
\end{eqnarray}
The ciclicity condition is assumed, namely for any $V_i\equiv B_i, C_i, E_i, F_i$ we get
$V_{i+2n}=V_i$. \par In the limit $n\rightarrow\infty$ the discretized coordinate $i$ describes the
compactified coordinate $\tau\in [0, 2\pi R]$, through
\begin{eqnarray}
V_i &\sim& V(\tau)\nonumber\\
\frac{V_{i+1}-V_i}{\Delta} &\sim& \frac{\partial}{\partial\tau}V, \quad with \quad \Delta = \frac{ 2\pi R}{2 n}\nonumber
\end{eqnarray}
By implementing the $N=1$ Lax pair system of \cite{tz} discussed in the previous Section, we get the following systems of super-Toda equations:\\
{\em ia}) case, given by
\begin{eqnarray}
D_+D_-B_i &=& \exp(E_i)-\exp(E_{i-1}),\nonumber\\
D_+D_-C_i &=& \exp(E_i)+\exp(E_{i+1}),\nonumber\\
D_+D_-E_i&=& \exp(-B_{i+1})-\exp(-B_i),\nonumber\\
D_+D_-F_i&=&\exp (-B_{i+1})+\exp(-B_i)
\end{eqnarray}
By conveniently reabsorbing the $\Delta$ factor in the redefinition of $\tau$, we get in the continuum limit
\begin{eqnarray}
D_+D_- B &=& \frac{\partial}{\partial\tau} \exp(E),\nonumber\\
D_+D_- E &=& \frac{\partial}{\partial\tau} \exp(-B)
\end{eqnarray}
together with
\begin{eqnarray}
D_+D_- C &=& 2\exp (E),\nonumber\\
D_+D_- F &=& 2\exp (-B)
\end{eqnarray}
It corresponds to two coupled Heavenly Equations with the ``wrong" sign, plus two infinite sets (labeled by $\tau$) of
equations of
Liouville-type in the heavenly background.\\
{\em ib}) case, given by
\begin{eqnarray}
D_+D_-B_i &=& \exp(F_i)-\exp(F_{i-1}),\nonumber\\
D_+D_-C_i &=& \exp(F_i)+\exp(F_{i-1}),\nonumber\\
D_+D_-E_i&=& \exp(C_{i+1})-\exp(C_i),\nonumber\\
D_+D_-F_i&=&\exp (C_{i+1})+\exp(C_i)
\end{eqnarray}
The continuum limit gives us
\begin{eqnarray}
D_+D_- C &=& 2 \exp(F),\nonumber\\
D_+D_- F &=& 2\exp(C)
\end{eqnarray}
together with
\begin{eqnarray}
D_+D_- B &=& \frac{\partial}{\partial\tau} \exp (F),\nonumber\\
D_+D_- E &=& \frac{\partial}{\partial\tau} \exp (C)
\end{eqnarray}
It corresponds to two infinite ($\tau$-dependent) series of coupled Liouville Equations with the ``good" sign.
$D=F$ is a consistent $N=1$ reduction, while the complexified system admits an $N=2$ supersymmetry and $D=F^\ast$ 
is a consistent $N=2$ reduction. The remining two equations are of 
Heavenly-type in the Liouville background.\\
{\em iia}) case, given by
\begin{eqnarray}
D_+D_-B_i &=& \exp(-E_i)-\exp(-E_{i-1}),\nonumber\\
D_+D_-C_i &=& \exp(-E_i)+\exp(-E_{i-1}),\nonumber\\
D_+D_-E_i&=& \exp(-B_{i+1})-\exp(-B_i),\nonumber\\
D_+D_-F_i&=&\exp (-B_{i+1})+\exp(-B_i)
\end{eqnarray}
The continuum limit gives us
\begin{eqnarray}
D_+D_- B &=& \frac{\partial}{\partial\tau} \exp(-E),\nonumber\\
D_+D_- E &=& \frac{\partial}{\partial\tau}\exp(-B)
\end{eqnarray}
together with
\begin{eqnarray}
D_+D_- C &=&  2\exp (-E),\nonumber\\
D_+D_- F &=& 2\exp (-B)
\end{eqnarray}
It corresponds to two coupled Heavenly Equations with the ``good" sign.
$B=E$ is a consistent $N=1$ reduction, while the complexified system admits $N=2$ supersymmetry
and $B=E^\ast$ is a consistent $N=2$ reduction. This is the $N=2$ Heavenly equation system introduced in
\cite{pt}. The two remaining equations are of
Liouville-type in the Heavenly background.\\
{\em iib}) case, given by
\begin{eqnarray}
D_+D_-B_i &=& \exp(F_i)-\exp(F_{i-1}),\nonumber\\
D_+D_-C_i &=& \exp(F_i)+\exp(F_{i-1}),\nonumber\\
D_+D_-E_i&=& \exp(-C_{i+1})-\exp(-C_i),\nonumber\\
D_+D_-F_i&=&\exp (-C_{i+1})+\exp(-C_i)
\end{eqnarray}
The continuum limit gives us
\begin{eqnarray}
D_+D_- C &=& 2 \exp(F),\nonumber\\
D_+D_- F &=& 2\exp(-C)
\end{eqnarray}
together with
\begin{eqnarray}
D_+D_- B &=& \frac{\partial}{\partial\tau} \exp (F),\nonumber\\
D_+D_- E &=& \frac{\partial}{\partial\tau} \exp (-C)
\end{eqnarray}
It corresponds to two infinite ($\tau$-dependent) sets of coupled Liouville Equations with the ``wrong" sign
(the system is $N=1$ supersymmetric only and there are no consistent reduction), plus two equations of
Heavenly-type in the Liouville background.

\section{The ${\bf S}^{1}/{\bf Z}_2$ coset construction.}

Let us analyze specifically the
closed reduced system for the {\em iia}) case. We have the system of super-Toda equations
\begin{eqnarray}
D_+D_- B_i &=& \exp(-E_i)-\exp(-E_{i-1}),\nonumber\\
D_+D_- E_i &=& \exp(-B_{i+1})-\exp(-B_i)
\end{eqnarray}
It is convenient to set
\begin{eqnarray}
B_i&=& A_{2i},\nonumber\\
E_i &=& A_{2i+1}\end{eqnarray}
in order to re-express the above system as 
\begin{eqnarray}
D_+D_- A_{2i} &=& \exp(-A_{2i+1})-\exp(-A_{2i-1}),\nonumber\\
D_+D_- A_{2i+1}&=&\exp(-A_{2i+2})-\exp(-A_{2i}).
\end{eqnarray}
We can write, compactly,
\begin{eqnarray}\label{Asystem}
D_+D_- A_j &=& \exp(-A_{j+1})-\exp(-A_{j-1}),
\end{eqnarray}
where $j$ ($A_{j+4n}=A_j$, see the previous section discussion) represents the discretization of the angular coordinate
$\theta_j =\frac{2\pi j}{4 n}$.
In the continuum limit this system of equations reads as follows
\begin{eqnarray}
D_+D_- A&=& \frac{\partial}{\partial\tau} \exp(-A).
\end{eqnarray}
The (\ref{Asystem}) discrete system, besides the ${\bf S}^1$ continuum limit (``$j$"
corresponds to the point labeled by $x_j= R cos\theta_j$, $y_j=Rsin\theta_j$) provides a discretization of the ${\bf I} =[-R,R]$ 
line since one can consistently project the associated $A_j$ (super)fields onto the circle diameter, 
by setting $A(x_j,y_j)\equiv A(x_j,-y_j)$ for $(x_j, y_j)\mapsto x_j$.
Indeed, it is easily proven that the constraint
\begin{eqnarray}
A_{4n-j} (x_+, \theta_+, x_-, \theta_-) &=& A_j (x_+, \theta_+, -x_-, -\theta_-).
\end{eqnarray}
is consistent w.r.t. the (\ref{Asystem}) equations of motion. Please notice the change in sign
in front of $x_-$, $\theta_-$, introduced 
in order to compensate the presence of a minus sign in the r.h.s. of (\ref{Asystem}).\par
The above constrained system corresponds to a discretized version of the supersymmetric heavenly equation
in the interval ${\bf I}$, with the superfields labeled by $A_0$ and $A_{2n}$ corresponding to the initial and final
position.

\section{The $N=2,4$ super-hydrodynamical reductions in $(1+1)$ dimensions}

The non-linear systems in $(1+2)$ dimensions derived in the previous section can be dimensionally
reduced by setting $x_+=x_-=x$. These reduced $(1+1)$-dimensional systems inherit the integrability properties
of their higher-dimensional analogs. In \cite{pt} it was shown that these types of non-linear equations can be recasted
as super-hydrodynamical types of equations, generalizing the bosonic hydrodynamical results of \cite{fp}. 
\par
The $(1+1)$-dimensionally reduced equations are supersymmetric, with twice as many supersymmetries as their corresponding
$(1+2)$-dimensional systems ($N=1\rightarrow N=2$, $N=2\rightarrow N=4$). 
The $N=4$ systems are obtained in correspondence with the (complexified) {\em ib}) and {\em iia}) cases. The
$N=4$ generators, satisfying 
\begin{eqnarray}
\{ Q_i,Q_j\} &=& -2 \delta_{ij}\partial_x
\end{eqnarray}
are explicitly given by
\begin{eqnarray}
Q_1 &=& \frac{\partial}{\partial\theta_+}
-{\theta}_+ \partial_x,\nonumber\\
Q_2 &=& \frac{\partial}{\partial\theta_-}
-{\theta}_- \partial_x,\nonumber\\
Q_3 &=& i( \frac{\partial}{\partial\theta_+}
+{\theta}_+ \partial_x),\nonumber\\
Q_4 &=& i(\frac{\partial}{\partial\theta_-}
+{\theta}_- \partial_x).
\end{eqnarray}
It is convenient to explicitly present one of these systems 
(the {\em iia})) in components. 
\par
Let us have
\begin{eqnarray}
B&=& b+\theta_+\psi_++\theta_-\psi_-+\theta_+\theta_-a,
\nonumber\\
C&=& c+\theta_+\xi_++\theta_-\xi_-+\theta_+\theta_-d,
\end{eqnarray}
with complex, $a,b,c,d$ bosonic and $\psi_\pm, \xi_\pm$ fermionic fields, all dependent on
$x$ and $\tau$. We further set the consistent constraints $E=B^\ast$, $F=C^\ast$. The
$(1+1)$-dimensionally reduced, type $iia$) system is equivalent to the equations
\begin{eqnarray}
a &=&-\frac{\partial }{\partial \tau }e^{-b^{\ast }},  \nonumber  \label{n4}
\\
\partial _{x}\psi _{\pm } &=&\pm \frac{\partial }{\partial \tau }%
(e^{-b^{\ast }}{\psi _{\mp }}^{\ast }),  \nonumber \\
{\partial _{x}}^{2}b &=&-\frac{\partial }{\partial \tau }[e^{-b^{\ast
}}(a^{\ast }+{\psi _{+}}^{\ast }{\psi _{-}}^{\ast })],
\end{eqnarray}
together with the background equations 
\begin{eqnarray}
d &=&-2e^{-b^{\ast }},  \nonumber \\
\partial _{x}\xi _{\pm } &=&\pm 2(e^{-b^{\ast }}{\psi _{\mp }}^{\ast }), 
\nonumber \\
{\partial _{x}}^{2}c &=&2e^{-b^{\ast }}\left( \frac{\partial }{\partial \tau}e^{-b}-{\psi _{+}}^{\ast }{\psi _{-}}^{\ast }\right) 
\end{eqnarray}
($a$ and $d$ are the auxiliary fields).

\section{Conclusions}

In this work we constructed a set of $N=1$ and $N=2$ supersymmetric ($2+1$)-dimensional integrable systems
arsing as continuum limit of super-Toda models obtained from the 
affine $sl(2n|2n)^{(1)}$ superalgebras series, in connection with different presentations for the
associated Cartan matrices. The equations, of heavenly type, were derived by using a manifest $N=1$ superLax formalism
and
involved in all cases under considerations two coupled superfields and two extra superfields in the
background produced by the two previous ones. These systems can be easily extended, {\em still preserving integrability},
by making the associated superfields valued not only on the real and complex numbers, but, e.g., on quaternions
as well. As an example, the $N=2$ {\em iia}) coupled system of equations, extended to the quaternionic
superfields $B= B_0+\sum_{i=1,2,3}e_iB_i$, $E= E_0+\sum_{i=1,2,3}e_iE_i$ (with $e_i$'s the three imaginary quaternions 
satisfying $e_ie_j=-\delta_{ij}+\epsilon_{ijk}e_k$) reads as follows, in terms of the real superfields
\begin{eqnarray}
D_{+}D_{-}B_{0} &=&\frac{\partial }{\partial \tau }\left( \exp (-E_{0})\cos (%
{\cal E})\right) ,  \nonumber \\
D_{+}D_{-}B_{i} &=&-\frac{\partial }{\partial \tau }(\exp (-E_{0})\frac{\sin
({\cal E})}{{\cal E}}E_{i}),  \nonumber \\
D_{+}D_{-}E_{0} &=&\frac{\partial }{\partial \tau }\left( \exp (-B_{0})\cos (%
{\cal B})\right) ,  \nonumber \\
D_{+}D_{-}E_{i} &=&-\frac{\partial }{\partial \tau }\left( \exp (-B_{0})%
\frac{\sin {\cal B}}{{\cal B}}B_{i}\right), 
\end{eqnarray}
where ${\cal E=}\sqrt{E_{1}^{2}+E_{2}^{2}+E_{3}^{2}}$ and ${\cal B=}\sqrt{%
B_{1}^{2}+B_{2}^{2}+B_{3}^{2}}$.\par
These equations inherit the integrability property from the corresponding (super)Lax presentation which can automatically
accommodate quaternionic-valued superfields.
It is worth noticing that, despite the fact that the bosonic sector of $sl(2n|2n)^{(1)}$ is quaternionic (see \cite{sstv}), 
the above system of equations is only $N=2$ supersymmetric; therefore the {\em quaternionization} does not induce any
hidden $N=4$ supersymmetry, while the {\em complexification}, in those selected cases that have been here discussed,
can induce a hidden $N=2$.\par  
An extended $N=4$ supersymmetry is only recovered when performing the \par $(1+2)\mapsto (1+1)$dimensional reduction,
see (\ref{n4}).\par
Perhaps one of the most challenging open problems concerns the existence of integrable $N>2$ supersymmetric extensions
of the heavenly equation directly in $(1+2)$-dimensions. It is unclear whether, let's say an $N=4$, integrable heavenly
equation indeed exists. This problem seems to be linked with the possibility of
reformulating the $N=4$ version of the Liouville equation (constructed \cite{ik} in the eighties in terms of quaternionic harmonic superfields) as a super-Toda system derived from a super-Lax based on a given, correctly identified,
Cartan matrix superalgebra. To our knowledge, this is still an unsolved problem which deserves being addressed.

\vskip1cm
{\large{\bf Acknowledgments.}}
~{\quad}\\{\quad}\par One of us (F.T.) is grateful for the hospitality at
the Institute of Theoretical Physics of the
University of Wroc{\l}aw, where this work was initiated.

\end{document}